\documentclass[prb,twocolumn,superscriptaddress,citeautoscript,showpacs,amsart]{revtex4-1}

\usepackage{graphicx}
\usepackage{multirow}
\usepackage{color}
\usepackage{bm}
\usepackage{times}
\usepackage{amsmath,bm,amsfonts}
\usepackage{dcolumn}
\usepackage{graphicx}
\usepackage{latexsym}
\usepackage{cancel}
\usepackage{hyperref}
\usepackage{physics}
\usepackage{ulem} 
\usepackage{braket}

\def\ket#1{|#1\rangle }
\def\bra#1{\langle #1 |}
\def\n{\nonumber \\ }

\newcommand{\ma}{\sigma}

\newcommand{\dt}{\delta}
\newcommand{\ep}{\epsilon}
\newcommand{\la}{\lambda}
\newcommand{\f}{\frac}

\newcommand{\Dt}{\Delta}
\newcommand{\dg}{\dagger}
\newcommand{\kb}{\vec{k}}
\newcommand{\Sb}{\vec{S}}

\newcommand{\A}{\alpha}
\newcommand{\B}{\beta}

\newcommand{\up}{\uparrow}
\newcommand{\dw}{\downarrow}
\newcommand{\al}[1]{\langle #1 \rangle}
\newcommand{\mtx}[1]{\left(\begin{matrix}#1\end{matrix}\right)}
\newcommand{\gm}{\gamma}

\newcommand{\de}{\tilde}
\newcommand{\db}{\vec{d}}

\newcommand{\Db}{\vec{D}}

\newcommand{\nm}{\nonumber \\ &}
\newcommand{\alg}[1]{\begin{align}#1\end{align}}
\newcommand{\w}[1]{\omega_{#1}}
\newcommand{\nq}{\nonumber \\ =&}

\begin{document}

\title{Nonreciprocal transport in $U(1)$ gauge theory of high-$T_c$ cuprates}

\author{Taekoo \surname{Oh}}
\affiliation{RIKEN Center for Emergent Matter Science (CEMS), Wako, Saitama 351-0198, Japan}

\author{Naoto \surname{Nagaosa}}
\email{nagaosa@riken.jp}
\affiliation{RIKEN Center for Emergent Matter Science (CEMS), Wako, Saitama 351-0198, Japan}

\date{\today}

\begin{abstract}
The nature of the charge carriers in high-$T_c$ cuprates is an essential issue to reveal their novel physical properties and the mechanism of their superconductivity. 
However, the experimental probes and the theoretical analysis have been mostly restricted to the linear responses. 
On the other hand, recent observations of Rashba-type spin-orbit coupling (SOC) on the surface of high-$T_c$ cuprates imply the possible nonlinear and nonreciprocal transport phenomena under in-plane magnetic fields. 
In this paper, we study the nonreciprocal transport properties on the surface of cuprates by employing a $U(1)$ gauge theory framework, where the electrons are considered to be fractionalized. 
Our investigation highlights the intricate variations in nonreciprocal transport with respect to temperatures and dopings. 
First, it reveals contrasting behavior of nonreciprocity in each normal phase. Second, it discerns the tendencies between underdoped and overdoped superconducting states and their paraconductivity. 
The complex behaviors of nonreciprocal transport originate from the spinful and spinless nature of the charge carriers and their kinetic energy scales.
These findings pave a new avenue to explore the electronic states in high-$T_c$ cuprates in terms of nonreciprocal 
transport phenomena. 
\end{abstract}


\maketitle

\begin{figure}
    \centering
    \includegraphics[width=\columnwidth]{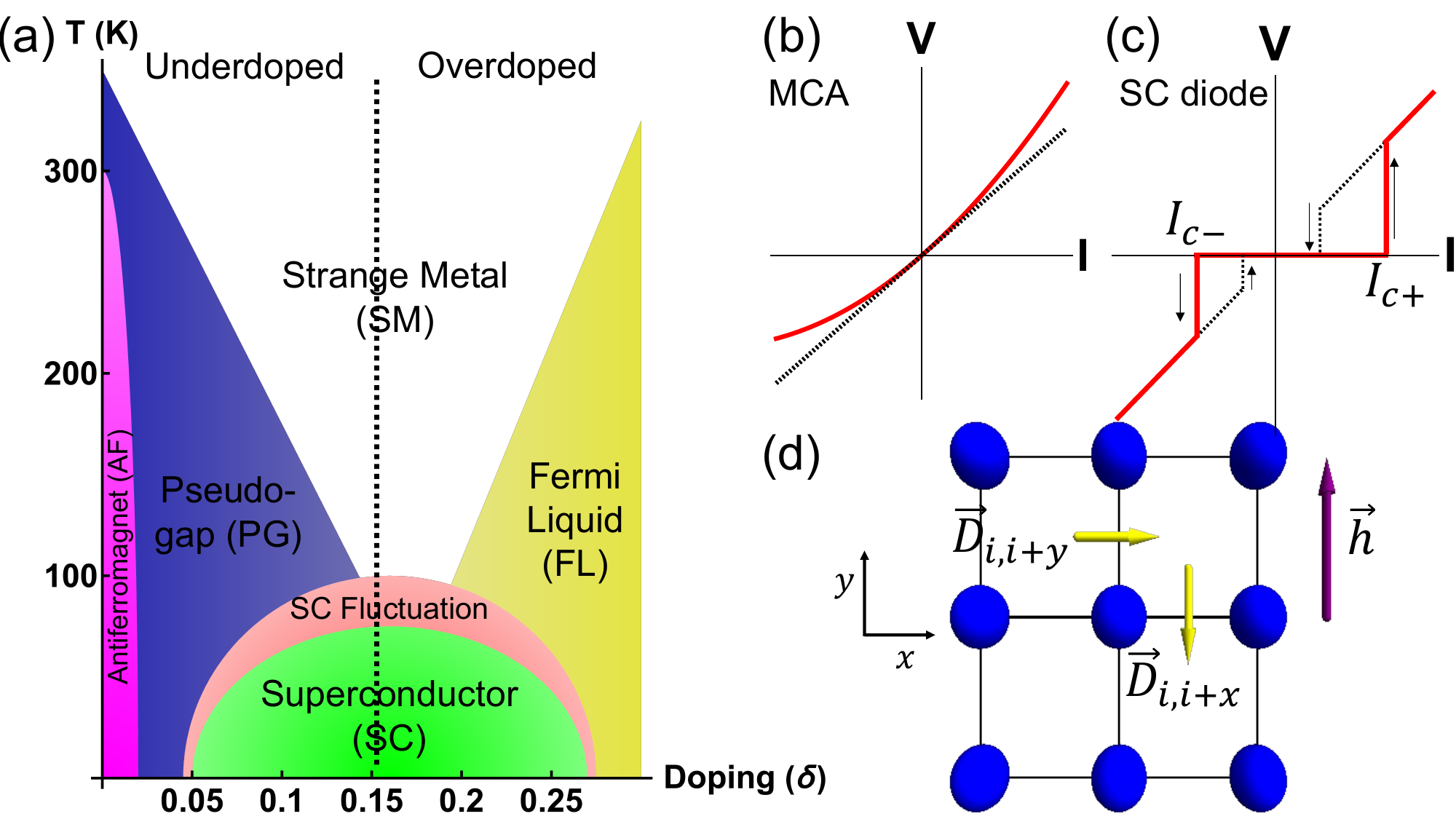}
    \caption{(a) The phase diagram of high-$T_c$ cuprates in $\delta$ and $T$ is divided into four regions. 
   The schematics of (b) MCA and (c) the SC diode effect. (d) The lattice structure of single CuO$_2$ 
   layer and directions of $\Db_{ij}\parallel \db_{ij}$ and $\vec{h}$.}
    \label{fig:1}
\end{figure}

{\it Introduction.|}
Several decades have passed since the groundbreaking discovery of high-$T_c$ superconductors in 
doped cuprates~\cite{bednorz1986possible}. 
The parent phase of undoped cuprates is the antiferromagnetic Mott insulator, and electron or hole doping 
leads to a complex phase diagram, encompassing the pseudogap, strange metal, Fermi liquid, 
and $d$-wave superconductor phases. [See Fig.~\ref{fig:1}(a).] Despite extensive research 
efforts~\cite{scalapino1995case,kastner1998magnetic,timusk1999pseudogap,orenstein2000advances,damascelli2003angle,
campuzano2004photoemission,lee2006doping,fischer2007scanning,alloul2009defects,armitage2010progress}, 
achieving a consensus on the origin of the phase diversity in cuprates remains elusive.

Notably, while cuprates are known for their weak spin-orbit coupling (SOC) compared with strong electronic 
correlations, recent experiments using spin-angle-resolved photoemission spectroscopy~\cite{gotlieb2018revealing} 
unveiled Rashba-type spin-orbit coupling on the surface of Bi$_2$Sr$_2$CaCu$_2$O$_{8+d}$. 
This arises from local inversion symmetry breaking at the surface of CuO$_2$ layers. Hence, the potential 
emergence of nonreciprocal transport in high-$T_c$ cuprates is anticipated when subjected to external magnetic 
fields, owing to both inversion ($\mathcal{P}$) and time-reversal ($\mathcal{T}$) symmetry breaking. 
Hence, the nonreciprocal transport could be the useful probe to investigate the physical properties of high-$T_c$ cuprates.

In metals, nonlinear nonreciprocal transport manifests through current-dependent resistance $R$~\cite{rikken2001electrical,Tokura2018,wakatsuki2017nonreciprocal,hoshino2018nonreciprocal}, expressed as:
\alg{
R = R_0(1+\gm h I).
}
Here, $R_0$ represents current-independent resistance, $\gamma$ denotes 
nonreciprocity, $h$ stands for the magnetic field, and $I$ corresponds to the current. This leads to asymmetric nonlinear $I-V$ 
curves in Fig.~\ref{fig:1}(b), a phenomenon known as magnetochiral anisotropy (MCA).

Conversely, in superconductors, nonreciprocal transport encompasses the nonlinear paraconductivity and the 
superconducting (SC) diode effect. Paraconductivity means a rapid increase in conductivity just above the 
transition temperature due to SC fluctuations, whose nonlinear part is also measured by 
$\gm$~\cite{wakatsuki2017nonreciprocal,wakatsuki2018nonreciprocal}. The SC diode effect measures the 
disparity in critical currents depending on the direction of the current flow, quantified by the quality factor $Q$~\cite{cui2019transport,ando2020observation,itahashi2020nonreciprocal,schumann2020possible,miyasaka2021observation,kawarazaki2022magnetic,narita2022field,masuko2022nonreciprocal,lin2022zero,scammell2022theory,he2022phenomenological,du2023superconducting,nadeem2023superconducting}, 
\alg{
Q = \frac{I_{c+} - I_{c-}}{(I_{c+}+I_{c-})/2}.\label{eq:2}
}
Here, $I_{c\pm}$ represent the critical currents for positive and negative current directions in Fig.~\ref{fig:1}(c), respectively.

In this Letter, we systematically explore the surface nonreciprocal transport across the complete phase diagram by 
employing the Slave boson $U(1)$ gauge theory of cuprates, where the electrons are considered be 
fractionalized into particles, i.e., fermionic spinons and bosonic holons. 
Upon the introduction of a magnetic field ($h$) into the system, 
we find the rich and nontrivial behaviors of the nonreciprocal transports in the plane of 
temperature ($T$) and doping ($\delta$). 
The phase diagram is partitioned into three regions: (I) the normal phases, including antiferromagnet (AF), pseudogap (PG), strange metal (SM), and Fermi liquid (FL) phases observed at high temperatures ($T\gg T_c$); (II) the phase marked by superconducting (SC) 
fluctuations at temperatures around $T\gtrsim T_c$; and (III) the $d$-wave superconducting 
phase existing at $T<T_c$ [refer to Fig.~\ref{fig:1}(a)].
In the AF, FL, and overdoped SC phases, the carriers are predominantly electrons, while in PG, SM, and underdoped SC phases, the carriers manifest as bosonic holons~\cite{lee1992gauge,nagaosa1992ginzburg}.

Each phase of normal states shows contrasting behavior of nonreciprocal transport, as for magnitude and dependence on $T$ and $\dt$.
Moreover, in both the SC fluctuation and $d$-wave SC phases, we find the significant increase of both $\gamma$ for paraconductivity and the quality factor $Q$ for the SC diode effect in the overdoped regions.
Based on the observations, we provide a global picture for nonreciprocal transport as well as practical predictions.

These observations are elucidated by the distinct characteristics of the carrier type and its kinetic energy.
The escalation of kinetic energy amplifies linear conductivity, diminishing nonreciprocal transport. 
Additionally, the absence of spin in holons results in an indirect Zeeman effect, and the symmetry breaking is weaker than electron case.
Accordingly, the magnitude of nonreciprocal transport in normal states is intricately linked to the kinetic energy of carriers, while its dependence on $T$ and $\dt$ predominantly comes from their symmetry breaking energy scales.
Similarly, the overdoped SC and its vicinity shows a sharp increase in nonreciprocal transports compared to the underdoped SC and its vicinity, owing to the carrier type and the fermion pairing energies.

{\it MCA in normal phases.|} 
To simulate the surface of doped cuprates under magnetic fields for normal phases at 
$T \gg T_c$, we employ the extended $t-J$ model $H_0 = H_t + H_{SO}+ H_Z + H_{S}$ and its associates in a 2D square lattice.
The components can be expressed as:
\alg{
H_t =& -t \sum_{\al{ij}} [c_{i\ma}^\dg c_{j\ma} + h.c.], \n
H_{SO} =& i\la_{SO} \sum_{\al{ij}} [c_{i\A}^\dg (\db_{ij}\cdot \vec{\ma}_{\A\B}) c_{j\B} + h.c.], \n
H_Z =& -\f{\vec{h}}{2}\cdot \sum_i c_{i\A}^\dg \vec{\ma}_{\A\B} c_{i\B}, \text{ and}\n
H_S =& \sum_{\al{ij}} [J \Sb_i \cdot \Sb_j + \Db_{ij} \cdot \Sb_i \times \Sb_j] - \vec{h}\cdot \sum_{i} \Sb_i. 
}
Here, $H_t$ represents the spin-independent hopping, $H_{SO}$ includes spin-dependent hopping due to SOC, 
and $H_Z$ accounts for the Zeeman effect of electrons. $H_S$ encompasses exchange, Dzyaloshinskii-Moriya (DM), 
and Zeeman interactions within the spin system. In Fig.~\ref{fig:1}(d), we illustrate the 2D square lattice with the directions of 
$\vec{D}_{ij} \parallel \vec{d}_{ij}$ and $\vec{h} \parallel \hat{y}$. The $U(1)$ Slave-boson theory is 
employed to solve $H_0$ for PG and SM phases. 
The model for AF and FL phases is derived from $H_0$. The spin exchanges are replaced with local electron spin couplings to magnetic order,
$H_{AF} = - 2u\sum_i \vec{m_i}\cdot \vec{j_i}$, where $\vec{m_i}$ is the local magnetic moment and $\vec{j}_i$ is the electron spin. 

To obtain the order parameters in the equilibrium state, we utilize self-consistent mean-field theory. The order parameters 
in the AF phase are staggered magnetic moments at A and B sublattices ($m_A$, $m_B$), while in the PG or SM phase, 
they include the spinonic singlet ($\chi$) and triplet ($\xi_x,\xi_y$) hoppings, singlet pairing ($\Delta$), 
and spinonic/holonic chemical potentials ($\mu_F/\mu_B$). Notably, $\vec{h}$ is instrumental 
in creating the asymmetry of the electronic band in AF and FL phases, while $\xi_x$ and $\xi_y$ contribute significantly to the asymmetry of the holonic band in PG and SM phases [see Supplementary Information (SI) for details].

\begin{figure}
    \centering
    \includegraphics[width=\columnwidth]{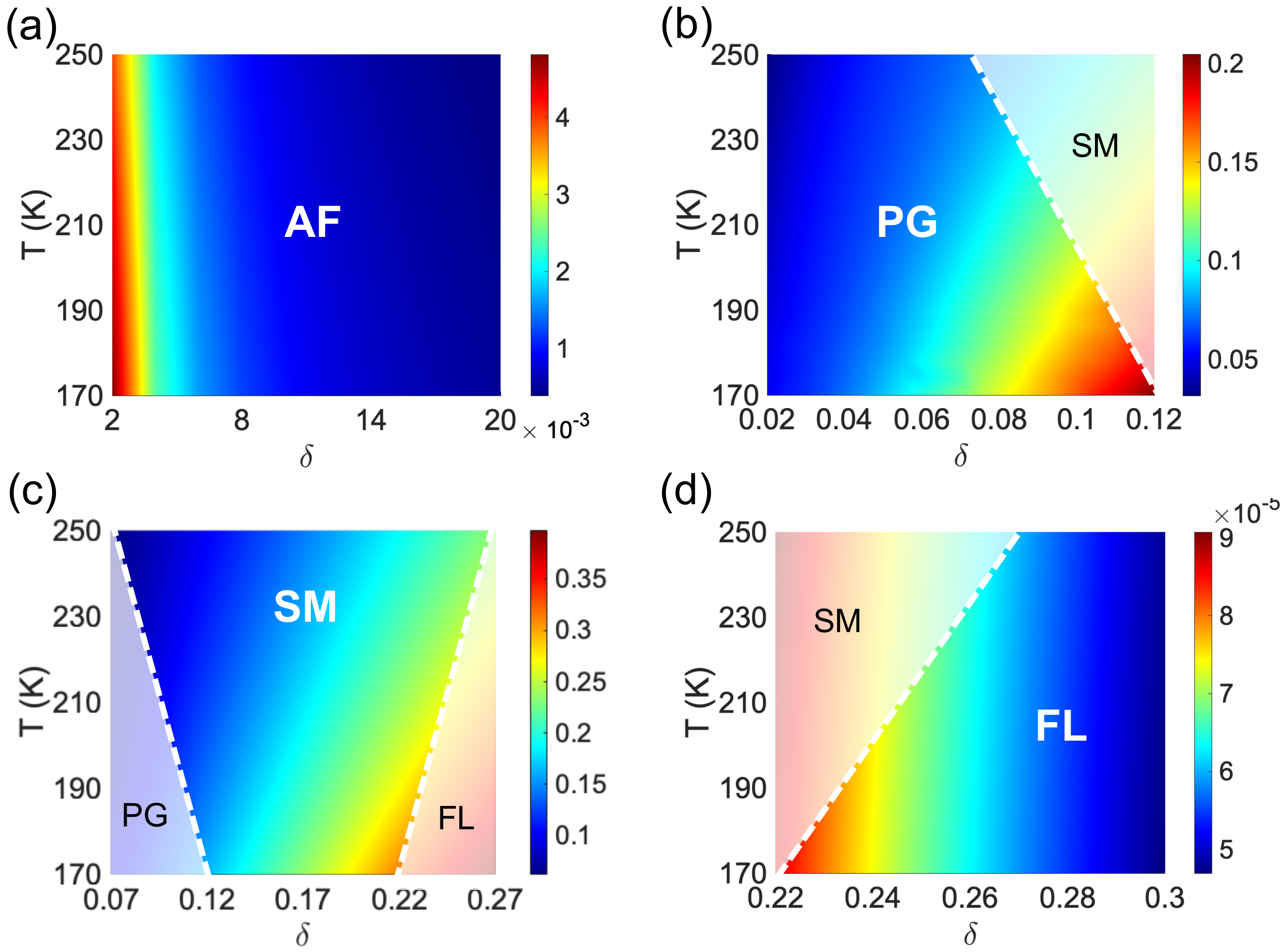}
    \caption{The nonreciprocity $\gm~ ($T$^{-1}$A$^{-1})$ in (a) AF ($\dt \in [0,0.02]$), (b) PG ($\dt \in [0.02,0.12]$), 
    (c) SM ($\dt \in [0.07,0.27]$), and (d) FL ($\dt \in [0.22, 0.3]$) phases.}
    \label{fig:2}
\end{figure}

For normal phases, we consider the longitudinal current up to the second order in an external electric field 
$\vec{\mathcal{E}}=Re[\vec{E}e^{i\omega t}]$ in a semiclassical manner. Accounting for SOC and $\vec{h}$, the magnetic space 
group reduces to $Pma'2'$, allowing for the second-order current along the $x$-axis. The current is defined as 
$J_x = Re[\ma_{1} E_x e^{i\w{}t} + \ma_{20}|E_x|^2 + \ma_{22}E_x^2 e^{2i\w{}t}]$, 
and the conductivities can be expressed as follows~\cite{nakai2019nonreciprocal}:
\alg{
\ma_1 =& - \frac{e^2\tau}{1+i\w{}\tau}\int d\epsilon ~\f{\partial f_0}{\partial\ep} \int_k v_x^2 \dt(\ep-\ep_k) , \label{eq:4}\\
\ma_{20} =& \f{e^3\tau^2}{4(1+i\w{}\tau)}\int d\ep ~\f{\partial^3 f_0}{\partial\ep^3} \int_k v_x^3 \Theta(\ep-\ep_k), \label{eq:5}  \\
\ma_{22} =& \f{\ma_{20}}{1+2i\w{}\tau}.
}
Here, $\int_k \equiv d^2k/(2\pi)^2$, $\tau$ represents the relaxation time, $f_0$ is the distribution function, 
$\epsilon_k$ is the energy dispersion, and $v_x$ is the $x$-direction of group velocity of the carriers. 
When $\omega = 0$, nonreciprocity can be approximated as 
$\gamma \approx -\frac{\ma_{22}}{\ma_1^2 W h}$, where $W$ represents the sample width, which is assumed to be $\sim 100~ \mu m$. Remarkably, $\gamma$ is independent of the relaxation time $\tau$.
Based on the measured values~\cite{zhang1988effective,kastner1998magnetic,lee2006doping,gotlieb2018revealing}, 
the parameters are fixed as $J=1$, $t=2$,  $|\mathbf{D}_{ij}| = 0.033$, $\lambda_{SO} = 0.067$, and $u=10$. The energy unit is $\sim 0.2~eV$, and the lattice constant is $4~\AA$. 
For the AF and FL phases, the electronic conductivity is computed while for the PG and SM phases, the holonic conductivity is computed.
The magnetic field $|\vec{h}| \sim 7 ~$T is fixed since $\gm$ is almost constant within $10~$T. 
The range of temperatures is from $170$ to $250~$K.
According to the phase diagram in Fig.~\ref{fig:1}(a), we set the doping ranges from $0$ to $0.02$ in AF phase, from $0.02$ to $0.12$ in PG phase, from $0.07$ to $0.27$ in SM phase, and from $0.22$ to $0.3$ in FL phase. The phase transition for AF, PG, and SM phases occurs at $T > 300~$K in the doping range.

\begin{figure}
    \centering
    \includegraphics[width=\columnwidth]{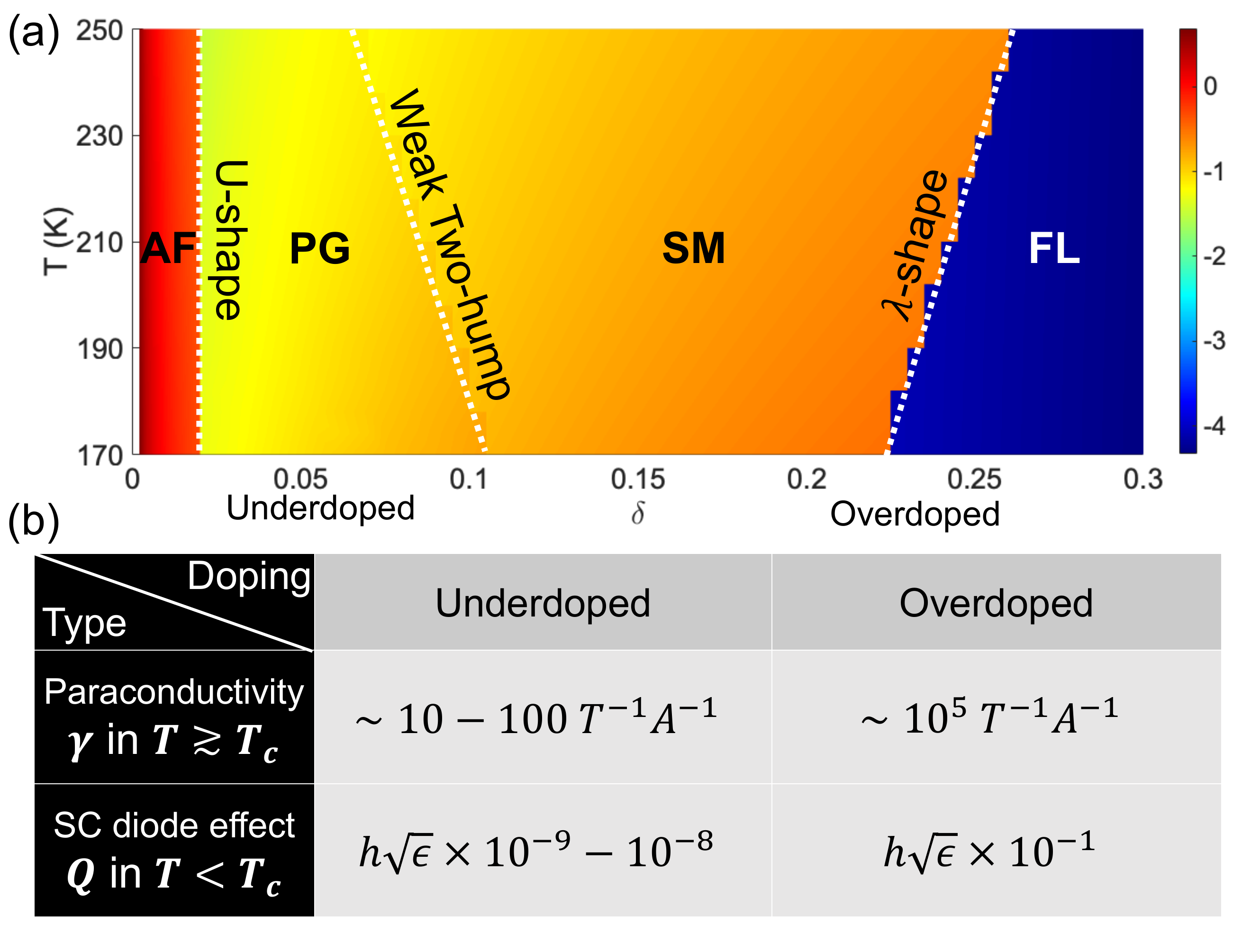}
    \caption{ (a) A guide map for nonreciprocity  $\log_{10} [\gm~($T$^{-1}$A$^{-1})]$ in temperatures and dopings for normal 
     phases. (b) The nonreciprocity $\gm$ of paraconductivity in $T\gtrsim T_c$ and the quality factor $Q$ of SC diode 
      effect in $T<T_c$. The unit of $h$ is tesla, and $\ep = |\ep'|$.}
    \label{fig:3}
\end{figure}

In Fig.~\ref{fig:2}, we present the change of $\gamma~$(T$^{-1}$A$^{-1}$) in temperatures and dopings for AF, PG, SM, and FL phases, respectively. 
We observe the followings. 
$\gm$ exhibits a sharp contrast for each phase in magnitude and the dependence on $T$ and $\dt$. 
Specifically, the magnitude of $\gm$ in PG and SM phases is much larger than that in FL phases, but much smaller than that in AF phases. 
As $T$ is lowered, $\gm$ grows in PG, SM, and AF phases while it slightly diminishes in FL phase.
As $\dt$ increases, $\gm$ grows in PG and SM phases while it diminishes in AF and FL phases.
Figure~\ref{fig:3}(a) shows the integrated diagram of $\log_{10}\gm$ relevant to the phase diagram in Fig.~\ref{fig:1}(a).

{\it Paraconductivity and SC diode effect.|} 
Meanwhile, to describe the paraconductivity and superconducting diode effect, we must formulate 
the Ginzburg-Landau (GL) free energy of SC phase. Let us assume that the GL free energy has the 
form: $F = \int_k (\eta(k) |\phi_k|^2 + \nu(k) |\phi_k|^4)$ with the order parameter $\phi_k$. 
From this free energy, we can compute the current due to paraconductivity above $T_c$ using the following 
equation~\cite{schmid1969diamagnetic,bennemann2008superconductivity,wakatsuki2018nonreciprocal},
\alg{
J_x = \f{2T_c}{V\Gamma}\sum_k \int_{-\infty}^t dt' j_x(t) \exp(-\f{2}{\Gamma} \int_{t'}^{t} dt'' \eta(t'')). \label{eq:7}
}
Here, $\Gamma$ represents the coefficient in the time-dependent GL equation, $V$ is the system 
volume, $\eta(t) = \eta(\vec{k}-2e\vec{E}t)$, and $j_x(t) \equiv -\partial\eta(t)/\partial A_x$~\cite{wakatsuki2018nonreciprocal}. 
Contrarily, the supercurrent below $T_c$ is computed using the derivative of the integral kernel of the 
free energy with respect to momentum, denoted as $J(k) = -2e[\eta'(k) |\phi_{k0}|^2 + \nu'(k)|\phi_{k0}|^4]$~\cite{he2022phenomenological}.
Here, $\phi_{k0}$ is the order parameter at the momentum $k$ corresponding to the energy 
minimum~\cite{he2022phenomenological}. The critical currents $I_{c+}$ and $-I_{c-}$ can be achieved at 
the maximum of $J(k)$ for positive momentum and the minimum for negative momentum, respectively. 
The quality factor is then computed as per Eq.~\ref{eq:2}.

The free energy for underdoped and overdoped SC phases should be different due to their distinct carriers and 
order parameters. In underdoped region, the carriers are predominantly holons near $T_c$. 
Considering that the superconducting phase transition occurs due to holon condensation, the free energy is as follows~\cite{nagaosa1992ginzburg}:
\alg{
F_1 = \int_k (\A_1 + \w{k})|\phi_k|^2 + \f{\B_1}{2}|\phi_k|^4,
}
Here, $\A_1 = \A_{10} T_c \ep'$ and $\B_1$ are the conventional Ginzburg-Landau parameters. 
$\phi_k$ is the superconducting order parameter, $\epsilon' = (T-T_c)/T_c$, and 
$\w{k} \equiv a_1k^2 - b_1 (k_x - k_x^3/6)$ is the holonic energy up to third order of momentum. 
By applying proper algebra, we find that the nonreciprocity of paraconductivity is 
$\gamma \sim 10-100~ $T$^{-1}$A$^{-1}$, and the quality factor of the SC diode effect is 
$Q\sim 10^{-9}h\sqrt{|\ep'|}$ [see SI for details].

On the other hand, in overdoped region, the carriers are mostly electrons near $T_c$. 
The superconducting phase transition here occurs due to the condensation of electron pairs. 
Furthermore in an $s$-wave SC, the SC diode effect can be present due to Rashba 
SOC~\cite{he2022phenomenological} while the nonlinear paraconductivity mainly arises from even and 
odd parity mixing by $\mathcal{P}$ symmetry breaking~\cite{wakatsuki2018nonreciprocal}. 
Although cuprates exhibit a $d$-wave SC, the underlying physics remains the same. 
Thus, we employ two distinct free energies $F_2$ and $F_3$ to represent $T \gtrsim T_c$ and $T < T_c$, respectively~\cite{mineev1999introduction,bauer2012non,wakatsuki2018nonreciprocal}:
\alg{
F_2 = & \int_k \sum_{AB} \psi_{Ak}^*[g^{-1}_{AB} + L_{Ak}\dt_{AB}]\psi_{Bk}, \label{eq:9}\\
F_3 = & \int_k \eta_3(k) |\psi_{k}|^2 + \nu_3(k) |\psi_{k}|^4. \label{eq:10}
}
Here, $\psi_k$ is the SC order parameter, $A$ and $B$ represent the matrix elements for even and odd parity order parameters, $g$ is the matrix for even and odd-parity electron interactions, $\eta_3(k)$ is a function of fourth order in momentum, and $L_{Ak}$ and $\nu_3(k)$ are the functions of second order in momentum.
The reciprocity of paraconductivity, $\gamma \sim 10^5~ T^{-1} A^{-1}$ 
and, the quality factor, $Q \sim 10^{-1}h\sqrt{|\ep'|}$, is obtained [see SI for details]. Compared to the underdoped 
case, the nonreciprocal transports sharply increase. The results are summarized in Fig.~\ref{fig:3}(b).

{\it Discussion.|} 
Nonreciprocal transport phenomena in high-$T_c$ cuprates are chiefly influenced by two factors: the kinetic energy and the type of carriers.
As the kinetic energy increases, the magnitude of linear conductivity becomes more prominent, so the nonreciprocal transport diminishes.
On the other hand, considering the spinlessness and indirect Zeeman effect of holons, nonreciprocal transport of holons will be small due to the weak symmetry breaking.
Specifically, in normal states, the magnitude of nonreciprocity $\gm$ is large in AF, intermediate in PG and SM, and small in FL, because the carriers gain more kinetic energy at a higher doping. 
Upon the kinetic energy, the drop of $\gm$ at the transition from AF to PG phases can be attributed to the fact that the carrier changes from electron to holon.
On the other hand, $\gm$ drops significantly at the transition from SM to FL despite electron carriers in FL phase, since the kinetic energy increases more rapidly.

Moreover, the disparity in energy scales of symmetry breaking of carriers dictates the dependence of $\gm$ on $T$ and $\dt$. 
In FL phase, the symmetry breaking energy scale of $h$ and $\la_{SO}$ remains constant with changes in $T$ and $\dt$, leading to a reduction in $\gm$ as $T$ decreases and $\dt$ increases due to kinetic energy. 
Similarly, in AF phase, upon $h$ and $\la_{SO}$, the symmetry breaking is governed by the magnetization which decreases as $T$ increases and $\dt$ decreases. 
This results in the enhancement in $\gm$ as $T$ increases, but $\gm$ still reduces in increasing $\dt$ because of the kinetic energy.
In contrast, in the PG and SM phases, the symmetry breaking is attributed to the energy scale of $\lambda_{SO}\xi_y$. This increases as $T$ decreases and $\dt$ increases, resulting in the enhancement of $\gm$ [see SI for details].

In the superconducting (SC) phase and its vicinity, where the system is regarded as a boson condensate, the energy scale of symmetry breaking assumes paramount importance. 
The breaking of $\mathcal{T}$ and $\mathcal{P}$ symmetries is attributed to $h$ and $\lambda_{SO}$ 
in electronic systems, whereas it is linked to $\lambda_{SO}\xi_y$ in holonic systems. Notably, since
$h \sim 1~meV$ and $\lambda_{SO}\sim 10~meV$ are much larger than $\lambda_{SO}\xi_y\sim 1~\mu eV$, the symmetry breaking is strong in the overdoped region.
Additionally, as the energy scale is introduced by the electron pairing $T_c \lesssim 70~K$, both $\gm$ for paraconductivity and the quality factor for the SC diode effect significantly enhance in the overdoped region.

In light of the theoretical analysis depicted in Figure \ref{fig:3}, we propose several predictions for experimental exploration. 
Discontinuities, suggested by the theoretical computations, might be replaced by a smooth but rapid change in reality, owing to the inherent approximations of the mean-field theory. 
Accordingly, we envisage distinctive nonmonotonous $\gm$ profiles, which begins with a $U$-shape in underdoped region and ends with a $\la$-shape in overdoped region as a function of $\dt$.
Moreover, the spinon pairing reduces the kinetic energy of holons, PG has a slightly larger nonreciprocity than SM. This might result in a weak two-hump shape for $\gm$ in $\dt$ on optimally doped region. 
Close to $T_c$, the surge in $\gm$ due to paraconductivity is expected to be more pronounced in the overdoped region compared to the underdoped region. 
In the SC phase, the rapid escalation of $Q$ in the SC diode effect driven by additional doping, is also foreseen.
Lastly, we address a caveat that our argument is not applicable to the bulk 3D system, since we only consider the 2D surface.

Our extensive theoretical examination underscores the intriguing aspects of surface nonreciprocal transport in diverse phases of high-$T_c$ cuprates and offers practical insights for future investigations. 
From the MCA in the AF phase to the nonlinear paraconductivity and SC diode effect in the SC phase, these findings illuminate the intricate behavior of high-$T_c$ cuprates. 
We are confident that these practical expectations will pave the way for pioneering discoveries in cuprates, serving as valuable tools in advancing the field of high-$T_c$ superconductors.

\begin{acknowledgments}
This work was supported by JST, CREST Grant Number JPMJCR1874, Japan. 
\end{acknowledgments}

\clearpage

\section*{Appendices to "Nonreciprocal transport in $U(1)$ gauge theory of high-$T_c$ cuprates"}

\appendix
\tableofcontents
\renewcommand{\thefigure}{S\arabic{figure}}
\renewcommand{\thetable}{S\arabic{table}}
\renewcommand{\arraystretch}{1.5}
\setcounter{figure}{0}
\setcounter{equation}{0}
\setlength{\tabcolsep}{3pt}

{The $t-J$ model and its associates}

\subsection{$t-J$ model}

The surface of doped cuprates is well described by the extended $t-J$ model, expressed as:
\alg{
	H = H_{S} + H_{t} + H_{SO} + H_Z,
}
where
\alg{
	H_S =& \sum_{\al{ij}} [J\Sb_i \cdot \Sb_j + \Db_{ij} \cdot \Sb_i \times \Sb_j] - \vec{h} \cdot \sum_i \Sb_i, \n
	H_t =& - t\sum_{\al{ij}} [c_{i\A}^\dg c_{j\A} + h.c.], \n
	H_{SO} =& i\la_{SO} \sum_{\al{ij}} [c_{i\A}^\dg (\db_{ij} \cdot \vec \ma_{\A\B}) c_{j\B} + h.c.],\n
	H_{Z} =& - \f{\vec{h}}{2} \cdot \sum_{i} c_{i\A}^\dg \vec{\ma}_{\A\B} c_{i\B}.
}
We define the vectors
\alg{
	&\Db_{ij} = -D\hat y, (j=i+x), D\hat x,(j=i+y); \n
	&\db_{ij} = - \hat y, (j=i+x), \hat x,(j=i+y); \text{ and} \n
	&\vec{h} = h \hat y.
}

We employ the Slave-boson U(1) theory as follows, 
\alg{
	c_{i\A}^\dg = f_{i\A}^\dg b_i, c_{i\A} = b_i^\dg f_{i\A}, \Sb_{i} = \f{1}{2} f_{i\A}^\dg \vec{\ma}_{\A\B} f_{i\B}.
}
Here, $f_{i\A}$ is fermionic, $b_i$ is bosonic.
Accordingly, we have
\alg{
	H_S =& -\de J \sum_{i} [A_{i,i+x}^\dg A_{i,i+x} + B_{i,i+x}^\dg B_{i,i+x} 
	\nm + A_{i,i+y}^\dg A_{i,i+y} + B_{i,i+y}^\dg B_{i,i+y}] 
	\nm - \de D \sum_i [- C_{y,i,i+x}^\dg B_{i,i+x} - B_{i,i+x}^\dg C_{y,i,i+x} \nm +  C_{x,i,i+y}^\dg B_{i,i+y} + B_{i,i+y}^\dg C_{x,i,i+y}] \nm 
	- \de h \sum_i (-i)(f_{i\up}^\dg f_{i\dw} - f_{i\dw}^\dg f_{i\up}) \nm + \sum_i \lambda_i (\sum_\A f_{i\A}^\dg f_{i\A} + b_i^\dg b_i - 1), \n
	H_t =& - t \sum_{i,\A} ( f_{i\A}^\dg b_i b_{i+x}^\dg f_{i+x\A} + f_{i\A}^\dg b_i b_{i+y}^\dg f_{i+y\A} + \text{h.c.}), \n
	H_{SO} =& \la_{SO} \sum_i( -  b_{i+x}^\dg b_i(f_{i\up}^\dg f_{i+x\dw} - f_{i\dw}^\dg f_{i+x\up}) \nm +   b_{i+y}^\dg b_i i(f_{i\up}^\dg f_{i+y\dw} + f_{i\dw}^\dg f_{i+y\up}) +  h.c.  ), \text{ and}\n
	H_Z =& -\de h \sum_i (-i)(f_{i\up}^\dg f_{i\up} - f_{i\dw}^\dg f_{i\dw}) b_{i}^\dg b_i.
}
Here, $\de J = J/4$, $\de D = D/4$, $\de h = h/2$, and
\alg{
	&A_{ij}^{\dg} = \ep_{\A\B} f_{i\A}^\dg f_{j\B}^\dg, A_{ij} = \ep_{\A\B} f_{j\B}f_{i\A} \n
	&B_{ij}^\dg = f_{i\A}^\dg f_{j\A}, B_{ij} = f_{j\A}^\dg f_{i\A}, \nm
	-iC_{x,ij}^\dg = f_{i\up}^\dg f_{j\dw} + f_{i\dw}^\dg f_{j\dw}, iC_{x,ij} = f_{j\up}^\dg f_{i\dw} + f_{j\dw}^\dg f_{i\dw}, \nm
	C_{y,ij}^\dg = (f_{i\up}^\dg f_{j\dw} - f_{i\dw}^\dg f_{j\up}), C_{y,ij} =  f_{j\dw}^\dg f_{i\up}- f_{j\up}^\dg f_{i\dw}.
}

Mean-field theory gives $\Dt = \al{A_{i,i+x}} = -\al{A_{i,i+y}}= \Dt^*, \chi = \al{B_{i,i+x}} = \chi^*, \xi_x = -i\al{C_{x,ij}^\dg} = \xi_x^*, \xi_y = -i\al{C_{y,ij}^\dg} = \xi_y^*$, and $m_y = -i\al{f_{i\up}^\dg f_{i\dw} - f_{i\dw}^\dg f_{i\up}}$. Here, $\Dt$ denotes the $d$-wave singlet spinon coupling, $\chi$ is the singlet spinon hopping, $\xi_x$ and $\xi_y$ are the triplet spinon hopping, and $m_y$ is the magnetization. We arrive at
\alg{
	H_F =& -\de J \sum_{i} [\Dt (A_{i,i+x}^\dg + A_{i,i+x} - A_{i,i+y}^\dg -  A_{i,i+y}) \nm + \chi (B_{i,i+x}^\dg + B_{i,i+x} + B_{i,i+y}^\dg + B_{i,i+y})] \nm 
	- i\de D \sum_i [-\chi \de C_{y,i,i+x}^\dg - \xi_y B_{i,i+x} + \xi_y B_{i,i+x}^\dg \nm + \chi \de C_{y,i,i+x} +  \chi \de C_{x,i,i+y}^\dg + \xi_x B_{i,i+y} - \xi_x B_{i,i+y}^\dg \nm - \chi \de C_{x,i,i+y}] 
	- \de h \sum_i (-i)(f_{i\up}^\dg f_{i\dw} - f_{i\dw}^\dg f_{i\up}) \nm + \mu_F \sum_i  (\sum_\A f_{i\A}^\dg f_{i\A} + x - 1), \n
	H_B =& -t \chi \sum_{i}( b_{i+x}^\dg b_i + b_{i+y}^\dg b_i + h.c.) \nm 
	+ i\la_{SO} \sum_i( - \xi_y b_{i+x}^\dg b_i  + \xi_x  b_{i+y}^\dg b_i   +  h.c.  ) 
	\nm - \de h  \sum_i m_y b_i^\dg b_i  + \mu_B \sum_i (b_i^\dg b_i - x).
}
where $\de C^\dg = -iC^\dg, \de C = iC$, and $x = \al{b_i^\dg b_i}$ is the hole doping. 

When we define $u_i = f_{i\up}$ and $d_i = f_{i\dw}$, the fermionic Hamiltonian is explicitly
\alg{
	H_F =& 
	-\de J \sum_{i} [\Dt (u_{i}^\dg d_{i+x}^\dg - d_{i}^\dg u_{i+x}^\dg + d_{i+x} u_{i}  
	- u_{i+x} d_{i} \nm
	- u_{i}^\dg d_{i+y}^\dg + d_{i}^\dg u_{i+y}^\dg - d_{i+y}u_{i} + u_{i+y} d_{i}) \nm + \chi (u_{i}^\dg u_{i+x} +d_{i}^\dg d_{i+x} + u_{i}^\dg u_{i+y} + d_{i}^\dg d_{i+y}+ h.c.)] \nm 
	+ \de D \sum_i \{\chi (u_{i}^\dg d_{i+x} - d_{i}^\dg u_{i+x} - u_{i+x}^\dg d_{i} + d_{i+x}^\dg u_{i}) \nm
	+ \xi_y (-i) (u_{i}^\dg u_{i+x} + d_{i}^\dg d_{i+x} - u_{i+x}^\dg u_{i} - d_{i+x}^\dg d_{i})) \nm
	+ (-i) [\chi (u_{i}^\dg d_{i+y}+ d_{i}^\dg u_{i+y} - u_{i+y}^\dg d_{i} - d_{i+y}^\dg u_{i} ) \nm 
	+ \xi_x (u_{i+y}^\dg u_{i} + d_{i+y}^\dg d_{i} - u_{i}^\dg u_{i+y} - d_{i}^\dg d_{i+y}) ]\} \nm
	- \de h \sum_i (-i)(u_{i}^\dg d_{i} - d_{i}^\dg u_{i}) \nm 
	+ \mu_F \sum_i  (u_i^\dg u_i + d_i^\dg d_i + x - 1),
}

\begin{figure}
	\centering
	\includegraphics[width=\columnwidth]{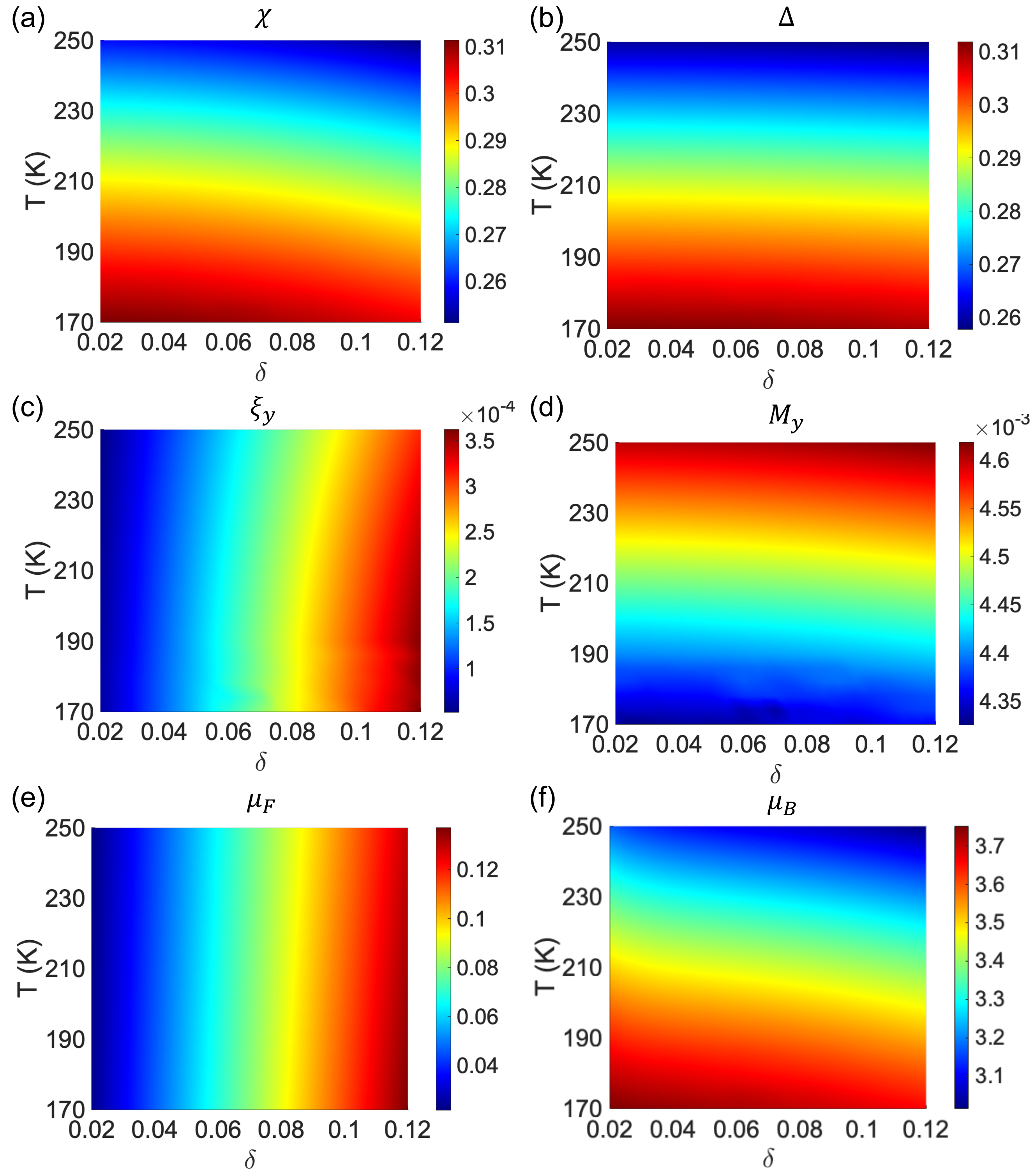}
	\caption{The order parameters of pseudogap phase. (a) The singlet hopping ($\chi$), (b) the singlet coupling ($\Dt$), (c) the triplet coupling ($\xi_y$), (d) the magnetization ($M_y$), (e) the fermionic chemical potential ($\mu_F$), and (f) the bosonic chemical potential ($\mu_B$). }
	\label{fig:S1}
\end{figure}

The Fourier transform of fermionic Hamiltonian is
\alg{
	H_F =&  -2\de J \sum_{k} [\Dt (u_{k}^\dg d_{-k}^\dg+ d_{-k} u_{k}) (\cos k_x - \cos k_y ) \nm
	+ \chi (u_{k}^\dg u_{k} +d_{k}^\dg d_{k})(\cos k_x + \cos k_y) ] \nm 
	+ 2\de D \sum_k \{\chi [i(u_{k}^\dg d_{k}  - d_{k}^\dg u_{k})\sin k_x \nm 
	+ (u_{k}^\dg d_{k}+ d_{k}^\dg u_{k}) \sin k_y] \nm
	+ (u_{k}^\dg u_{k}  + d_{k}^\dg d_{k}) (\xi_y \sin k_x - \xi_x \sin k_y)\} \nm
	- \de h \sum_k (-i)(u_{k}^\dg d_{k} - d_{k}^\dg u_{k}) \nm
	+ \mu_F \sum_k  (u_k^\dg u_k + d_k^\dg d_k).
}
In matrix form,
\alg{
	H_F = \f{1}{2} \sum_k \psi_k^\dg \mtx{ H_0(k) & \Dt_{gap}(k) \\ \Dt_{gap}^\dg(k) & -H_0(-k) } \psi_k,
}
where $\psi_k = [u_k,d_k,u_{-k}^\dg, d_{-k}^\dg]^T$,
\alg{
	H_0(k) =& [-2\de J \chi (\cos k_x + \cos k_y) + 2\de D (\xi_y \sin k_x \nm
	- \xi_x \sin k_y) + \mu_F] \tau_0  + 2\de D \chi \sin k_y \tau_1 \nm - (\de h +2\de D \chi \sin k_x) \tau_2,\n
	\Dt_{gap}(k) =& \mtx{ 0 & \Dt_d(k) \\ -\Dt_d(-k)& 0 },\text{ and}\n
	\Dt_d(k) =& -2\de J \Dt (\cos k_x - \cos k_y). \label{eq:S11}
}
Here, $\tau$ is the Pauli matrix representing the spin degrees of freedom. When $\ma$ are the Pauli matrices representing particle-hole degrees of freedom, we represent 
\alg{
	H(k) =& \f{1}{2} \{ [ -2\de J \chi (\cos k_x + \cos k_y) + \mu_F ] \tau_0 \ma_3 \nm
	+ [2\de D (\xi_y \sin k_x - \xi_x \sin k_y)] \tau_0 \nm 
	+ 2 \de D \chi (\sin k_y \tau_1 -  \sin k_x \tau_2) - \de h \tau_2 \ma_3 \nm
	- \Dt_d(k) \tau_2\ma_2 \}.
}

On the other hand, the Fourier transform for bosonic system gives
\alg{
	H_B =& - 2t \chi \sum_{k} b_k^\dg b_k (\cos k_x + \cos k_y)  \nm+ \la_{SO} \sum_k( - \xi_y b_{k}^\dg b_k \sin k_x   + \xi_x  b_{k}^\dg b_k \sin k_y)  ) \nm
	+ (- \de h m_y + \mu_B) \sum_i b_k^\dg b_k.
}
This is a single band system. The bosonic energy is
\alg{
	\ep_k =& -2t\chi(\cos k_x+\cos k_y) + \la_{SO} (\xi_x \sin k_y -\xi_y \sin k_x) \nm + \mu_B - \de h m_y. \label{eq:S14}
}
Notably, the symmetry is broken by $\la_{SO}\xi_x$ and $\la_{SO}\xi_y$ while the role of $h$ and $m_y$ is to shift the energy band. 

\begin{figure}
	\centering
	\includegraphics[width=\columnwidth]{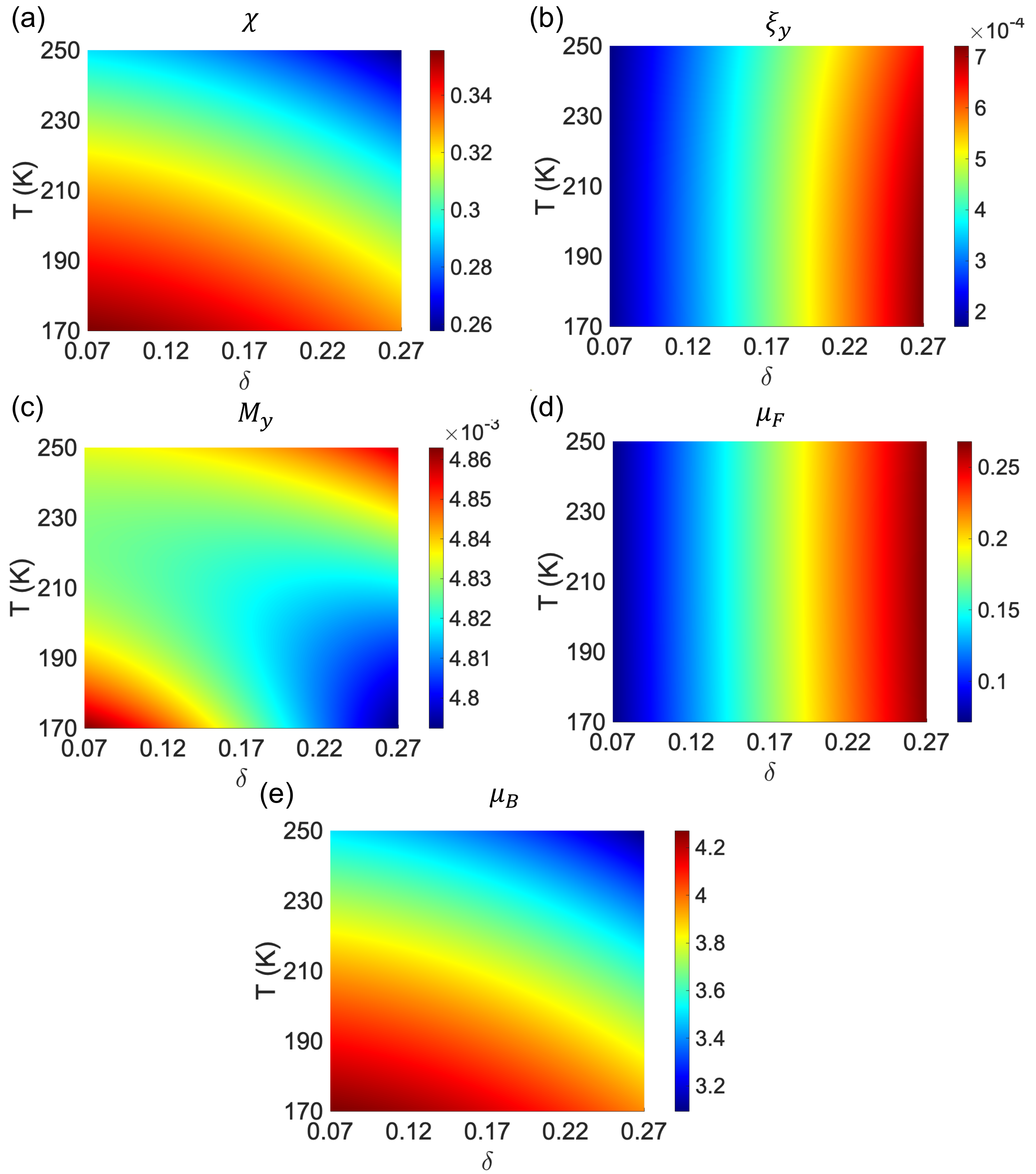}
	\caption{The order parameters of strange metal phase. (a) The singlet hopping ($\chi$), (b) the triplet coupling ($\xi_y$), (c) the magnetization ($M_y$), (d) the fermionic chemical potential ($\mu_F$), and (e) the bosonic chemical potential ($\mu_B$). }
	\label{fig:S2}
\end{figure}

The Green function for BdG fermionic Hamiltonian is
\alg{
	G(i\w{n},k) =& [i\w{n} - H(k)]^{-1} = \sum_a\f{\ket{a,k}\bra{a,k}}{i\w{n} - \w{k,a}} \nq \sum_a \f{U_{\A,a}U^\dg_{a,\B}}{i\w{n}-\w{k,a}}\ket{\A,k}\bra{\B,k},
}
where $\w{k,a}$ is the eigenenergy, $a=1,2,3,4$, and $U^\dg H U = D$. ($U_{\A,a} = \bra{\A,k}\ket{a,k},U^\dg_{a,\A} = \bra{a,k}\ket{\A,k}$)
The Green function can be represented as 
\alg{
	G_{\A\B}(\tau,k) =& -\al{T_\tau \psi_{k,\A}(\tau) \psi_{k,\B}^\dg } \nq \f{1}{\B} \sum_{i\w{n}} e^{-i\w{n}\tau} G_{\A\B}(i\w{n},k).
}
Then,
\alg{
	G_{\A\B}(0^-,k) =& \al{\psi_{k,\B}^\dg \psi_{k,\A} } = \f{1}{\B} \sum_{i\w{n}} e^{i\w{n}0^+} G_{\A\B}(i\w{n},k) \nq \oint \f{dz}{2\pi i} n_F(z) G_{\A\B}(z,k) \nq
	\oint \f{dz}{2\pi i} n_F(z) \sum_a \f{U_{\A,a}U^\dg_{a,\B}}{z-\w{k,a}}\nq
	\sum_a  U_{\A,a} n_F(\w{k,a}) U^\dg_{a,\B}
}
The self-consistent equations for mean-field parameters are following. When we denote $\int_k = \int \f{d^2 k}{(2\pi)^2}$, the singlet hopping is
\alg{
	\chi =& \f{1}{4N} \sum_i \al{B_{i,i+x} + B_{i,i+x}^\dg + B_{i,i+y} + B_{i,i+y}^\dg } \nq
	\f{1}{2N} \sum_k \al{u_k^\dg u_k + d_k^\dg d_k}(\cos k_x+\cos k_y) \nq
	\f{1}{4} \int_k \Tr[G(0^-,k)\tau_0 \ma_3](\cos k_x+\cos k_y).
}
Also, the singlet coupling is
\alg{
	\Dt =& \f{1}{4N} \sum_i \al{A_{i,i+x} +A_{i,i+x}^\dg  - A_{i,i+y}- A_{i,i+y}^\dg} \nq
	\f{1}{2N} \sum_k \al{u_{k}^\dg d_{-k}^\dg+ d_{-k} u_{k}} (\cos k_x - \cos k_y )
	\nq \f{1}{4} \int_k \Tr[G(0^-,k) (-\tau_2\ma_2) ] (\cos k_x - \cos k_y ).
}
In addition, the triplet hopping is 
\alg{
	\xi_x =& \f{1}{2N} \sum_i \al{\de C_{x,i,i+y} + \de C_{x,i,i+y}^\dg} \nq
	\f{1}{N} \sum_k \al{u_k^\dg d_k + d_k^\dg u_k} (\cos k_y) \nq
	\f{1}{2} \int_k \Tr[G(0^-,k) \tau_1 \ma_3] (\cos k_y).
}
Similarly, the other part is
\alg{
	\xi_y =& \f{1}{2N} \sum_i \al{\de C_{y,i,i+x} + \de C_{y,i,i+x}^\dg} \nq
	\f{1}{2N} \sum_i (-i) \al{u_i^\dg d_{i+x} - d_{i}^\dg u_{i+x} + u_{i+x}^\dg d_i - d_{i+x}^\dg u_i }\nq
	\f{1}{N} \sum_k (-i)\al{u_k^\dg d_k - d_k^\dg u_k}\cos k_x\nq
	\f{1}{2} \int_k \Tr[G(0^-,k) \tau_2 \ma_3] \cos k_x.
}
The magnetization is
\alg{
	m_x =& \f{1}{N} \sum_i \al{u_i^\dg d_i + d_i^\dg u_i}, \nq
	\f{1}{N} \sum_k \al{u_k^\dg d_k + d_k^\dg u_k},\nq
	\f{1}{2} \int_k\Tr[ G(0^-,k) \tau_1\ma_3],\n
	m_y =& \f{1}{N} \sum_i (-i) \al{u_i^\dg d_i - d_i^\dg u_i}, \nq
	\f{1}{N} \sum_k (-i) \al{u_k^\dg d_k - d_k^\dg u_k}, \nq
	\f{1}{2}\int_k \Tr[G(0^-,k)\tau_2\ma_3].
}
Finally, the number density constraint is
\alg{
	1-x =& \f{1}{N} \sum_i \al{u_i^\dg u_i + d_i^\dg d_i} \nq
	\f{1}{N} \sum_k \al{ u_k^\dg u_k + d_k^\dg d_k } \nq
	\f{1}{2}\int_k (\Tr[G(0^-,k) \tau_0\ma_3 ] + 2),\n
	-x =& \f{1}{2} \int_k \Tr[G(0^-,k) \tau_0\ma_3].
}

On the other hand, the bosonic Green function is given by
\alg{
	G_B(i\ep_n,k) = (i\ep_n - \ep_k)^{-1}.
}
Then,
\alg{
	G_B(0^-,k) =& \f{1}{\B}\sum_{i\ep_n} e^{i\ep_n 0^+}G_B(i\w{n},k) \nq
	-\oint \f{dz}{2\pi i} n_B(z) G_B(z,k),\nq
	-\oint \f{dz}{2\pi i} \f{n_B(z)}{z-\ep_k} = -n_B(\ep_k).
}
Thus,
\alg{
	x = \f{1}{N} \sum_i \al{b_i^\dg b_i} = \f{1}{N} \sum_k \al{b_k^\dg b_k} = \int_k n_B(\ep_k).
}

Typically, the energy is known to be $t\sim 400~$meV, $J\sim 200$~meV, $\la_{SO} \sim 0.03t$.
The parameters we use here are $t=3, J=1.5, \la_{SO} = 0.1, |\vec{D}| = 0.05$, and $h=0.005$ ($\sim 7-8$ T), where the unit is $\sim 0.12~eV$. 
The temperature varies from $0.100$ to $0.168$ ($\sim 150 - 250$ K). 
The doping range is $0.02 - 0.12$ for pseudogap phase, and $0.07-0.27$ for strange metal phase.
The order parameters for pseudogap and strange metal phases are in Figs.~\ref{fig:S1} and \ref{fig:S2}. Notably, for both cases, $M_x$ and $\xi_x$ are negligibly small due to the symmetry.

\begin{figure}
	\centering
	\includegraphics[width=\columnwidth]{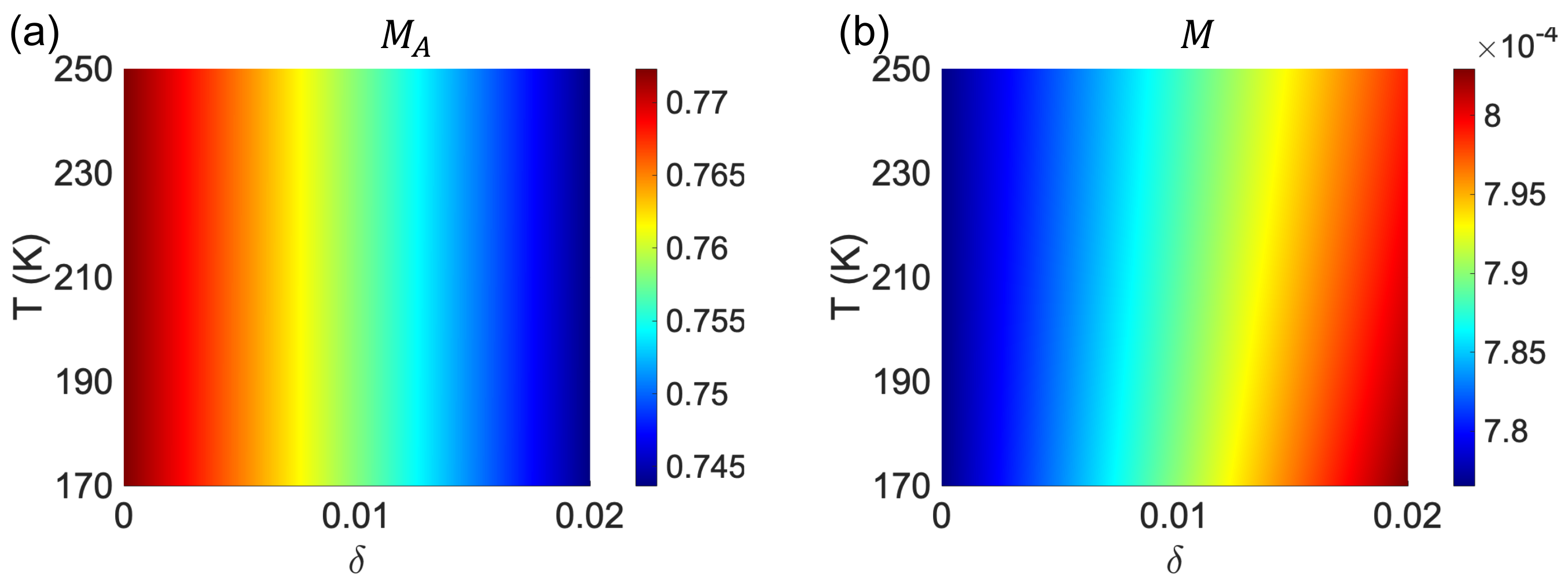}
	\caption{The order parameters of Hubbard model. (a) The antiferromagnetic order ($M_A$), (b) The ferromagnetic order ($M$). }
	\label{fig:S3}
\end{figure}

\subsection{The associated model derived from $t-J$ model}

The Fermi liquid and antiferromagnet phases are described by the following model derived from $t-J$ model:
\alg{
	H = H_{AF} + H_t + H_{SO} + H_z.
}
Since local electron spin couplings to the magnetic order becomes significant, we replace $H_S$ with 
\alg{
	H_{AF} = -2u \sum_i \vec{m}_i \cdot \vec{j}_i.
}
Here, $\vec{j}_i = \f{1}{2}c_{i\A}^\dg \vec{\ma}_{\A\B} c_{i\B}$ and $\vec{m}_i = \al{\vec{j}_i}$.

For Fermi liquid phase, we let $u=0$. For antiferromagnetic phase, we let $u=15$ ($u\sim2$~eV), and acquire $\al{\vec{j}_i}$ for two distinct sublattices. The doping range for Fermi liquid phase is $0.22 - 0.30$, while that for antiferromagnetic phase is $0 - 0.02$. For AF phase, the order parameters are $M_A = |(\vec{m}_1 - \vec{m}_2)/2|$ and $M = |(\vec{m}_1+\vec{m}_2)/2|$. 
Notably, $\mathcal{T}$ and $\mathcal{P}$ symmetries of the energy bands are broken by $\la_{SO}$, $h$, and $M$.
We represent the order parameters in Fig.~\ref{fig:S3}.

\begin{figure}
	\centering
	\includegraphics[width=\columnwidth]{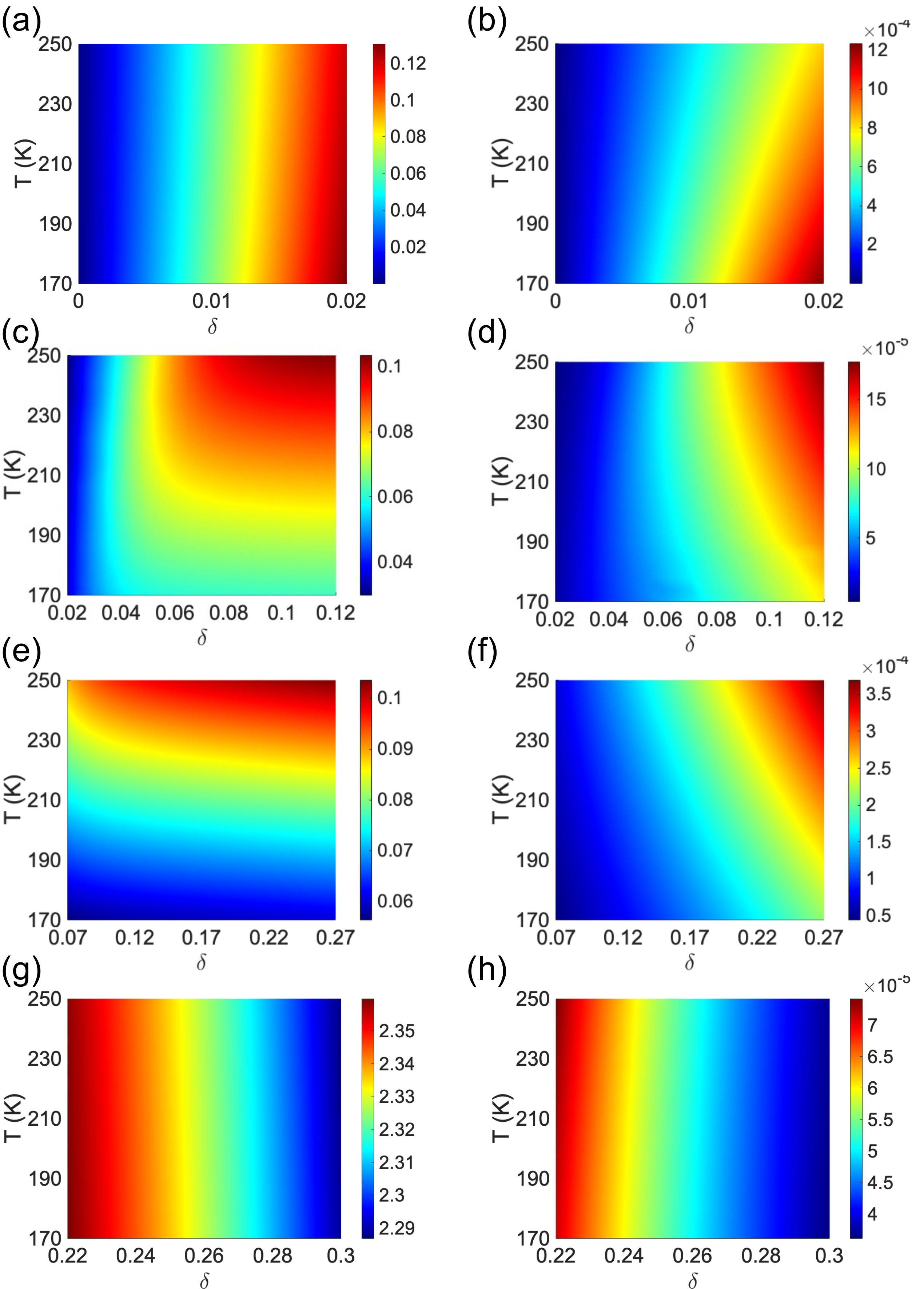}
	\caption{Linear (left panels; $V_x$) and second-order transport (right panels; $-C_x$) for (a-b) AF, (c-d) PG, (e-f) SM, (g-h) FL phases.}
	\label{fig:S4}
\end{figure}

\subsection{Nonreciprocal transport}

When the system is subject to the electric field $Re[\vec{E}e^{i\w{}t}]$, the current is represented as $J_x = Re[\ma_{1} E_x e^{i\w{}t} + \ma_{20}|E_x|^2 + \ma_{22}E_x^2 e^{2i\w{}t}]$. Then, in a semiclassical way, the conductivities are 
\alg{
	\ma_1(\w{}) =& \frac{e^2\tau}{1+i\w{}\tau}V_x, \n
	\ma_{20}(\w{}) =& \f{e^3\tau^2}{4(1+i\w{}\tau)}C_x,  \text{ and}\n
	\ma_{22}(\w{}) =& \f{\ma_{20}}{1+2i\w{}\tau},
}
where
\alg{
	V_x =& \int d\epsilon ~(-\f{\partial f_0}{\partial\ep}) \int_k v_x^2 \dt(\ep-\ep_k), \text{ and}\n
	C_x =& \int d\ep ~\f{\partial^3 f_0}{\partial\ep^3} \int_k v_x^3 \Theta(\ep-\ep_k).
}
Here, $\int_k = \int d^2k/(2\pi)^2$, $\ep_k$ is the energy dispersion, $v_x = \f{1}{\hbar}\f{\partial \epsilon_k}{\partial k_x}$ is the group velocity, $f_0$ is the distribution function, and $\tau$ is the lifetime of carriers. We use Fermi-Dirac distribution for AF and FL, but Bose-Einstein distribution for PG and SM phases. By letting $\hbar$ and the lattice constant to be a unit, we compute $V_x$ and $C_x$ for each phase and represent it in Fig.~\ref{fig:S4}. The nonreciprocity is defined as:
\alg{
	\gm \approx -\f{\ma_{22}(0)}{\ma_1^2(0) Wh} = -\f{C_x}{4eV_x^2Wh}.
}
Here, $W$ is the sample width which is assumed to be $\sim 100~\mu m .$

\clearpage

\section{The free energies of superconductors}

\subsection{Paraconductivity and SC diode effect}

We here describe the Ginzburg-Landau free energies and the nonreciprocal transports for superconductor phase.
The basic form of free energy is 
\alg{
	F = \int_k (\eta(k) |\phi_k|^2 + \nu(k) |\phi_k|^4).
}
Using this free energy, we obtain the nonreciprocity $\gm$ by computing the paraconductivity, and the quality factor for SC diode effect. The current from paraconductivity is defined as:
\alg{
	J_x = \f{2T_c}{V\Gamma}\sum_k \int_{-\infty}^t dt' j_x(t) \exp(-\f{2}{\Gamma} \int_{t'}^{t} dt'' \eta(t'')). \label{eq:S33} 
}
Here, $\Gamma$ is the coefficient in the time-dependent Ginzburg-Landau equation, $V$ is the volume of system, $\eta(t) = \eta(\vec{k}-2e\vec{E}t)$, and $j_x(t) \equiv -\partial\eta(t)/\partial A_x$. 

On the other hand, the supercurrent is computed by derivative of the integral kernel of the free energy with respect to momentum,
\alg{
	J(k) = -2e[\eta'(k) |\phi_{k0}|^2 + \nu'(k)|\phi_{k0}|^4].
}
Here, $\phi_{k0}$ is the order parameter at the energy minimum for momentum $k$. 
The critical currents $I_{c+}$ and $-I_{c-}$ are found at the maximum of $J(k)$ for positive momentum and the minimum for negative momentum, respectively. 
The quality factor is $Q = 2(I_{c+}-I_{c-})/(I_{c+}+I_{c-})$.

\subsection{Underdoped}

For underdoped case, the holonic transport is dominant. Because the holon is bosonic, the superconductor arises when the holonic condensation occurs. The order parameter for bose condensation is $\al{b}$, where $b$ is the boson operator. Accordingly, the free energy for holonic superconductor is simply
\alg{
	F_1 = \int_k (\A_1 + \ep_k) |\phi_k|^2 + \f{\B_1}{2} |\phi_k|^4.
}
Here, $\phi_k$ is the order parameter, $\A_1 = \A_0T_c \ep$ and $\B_1$ are conventional Ginzburg-Landau order parameters,  $\A_0 \sim H_{cB}^2/8\pi \sim J\dt^2$, $\ep = (T - T_c)/T_c$, and $\ep_k$ is the holonic energy.
From Eq.~\ref{eq:S14}, we approximate $\ep_k$ up to the third order of momentum near $\Gamma$ point, $\ep_k = a k^2 - \la (k_x - k_x^3/6)$, where $a\sim t\chi$, $\la \sim \la_{SO}\xi_y$.

Using Eq.~\ref{eq:S33}, the current from paraconductivity is given by
\alg{
	J = \f{e^2}{16\ep}E - \f{e^3\pi}{1536T_c\ep^2}\f{\la}{a}E^2.
}
Thus, the nonreciprocity is
\alg{
	\gm = \f{\pi\la}{6aheT_cW}.
}
The quality factor from SC diode effect, on the other hand,
is given by
\alg{
	Q = \f{\sqrt{-\A_1}\la}{3\sqrt{3}a^{3/2}}.
}

When we apply $t\sim 3, \chi \sim 1, \la \sim 0.1, \xi_y \sim 10^{-2} - 10^{-1}h, T_c \sim 10^{-2}$, $\A_0\sim J\dt^2,$ and $\dt \sim 0.1$, we got $\gm \sim 10- 100 ~T^{-1} A^{-1}$ and $Q \sim 10^{-9} - 10^{-8} h\sqrt{-\ep}$. 

\subsection{Overdoped}

For overdoped case, the electronic transport is dominant. The superconductor arises when the Cooper pair is formed. The order parameter here is $\al{cc}$, where $c$ is the fermionic operator. Hence, the free energy could be obtained by computing the bubble diagrams.

Here, while the SC diode effect comes from Rashba spin-orbit coupling, the nonlinear paraconductivity is attributed to the parity mixing by $\mathcal{P}$ symmetry breaking. Therefore, we employ two different free energies as for the temperature range. 

Basically, we suppose that the Hamiltonian in spinor basis is $H_0 = \sum_k c_{k\A}^\dg H_{k,\A\B} c_{k\B}$, where
\alg{
	H_k = \xi_k + a(k_x \ma_y - k_y \ma_x) - \vec{h}\cdot \vec{\ma}.
}
Here, $\xi_k = k^2/2m-\mu$. Comparing with Eq.~\ref{eq:S11}, we could let $m^{-1} \sim J\chi, a \sim D\chi$. The energy dispersion is
\alg{
	E_{kA} = \xi_k + A \sqrt{(ak_x-h)^2 + (ak_y)^2 },
}
where $A=\pm1$.

When $T\gtrsim T_c$, in order to consider the parity mixing, we include the following electron interactions in spinor basis:
\alg{
	H_{int}^g =& \f{1}{2\mathcal{V}} \sum_{\kb,\kb'} V^g(\kb,\kb') c_{k,\up}^\dg c_{-k,\dw}^\dg c_{-k',\dw} c_{k',\up} \n
	H_{int}^u =& \f{1}{2V}\sum_{\kb\kb'} V^u_{ij}(\kb,\kb') (i\ma_i\ma_2)_{\A\B} (i\ma_j\ma_2)^\dg_{\gm\dt} \n&\times c_{k\A}^\dg c_{-k\B}^\dg c_{-k'\gm}c_{-k'\dt},
}
Here, $V_g(\kb,\kb') = -V_g(k_x^2-k_y^2)(k_x'^2-k_y'^2)/2k^2k'^2$ and $V^u_{ij}(\kb,\kb') = V_u g_i(\kb) g_j(\kb')$, where $\vec{g}(\kb) = (-k_y,k_x)/k$. 

By transforming the Hamiltonians to the band representation and perform the calculation for bubble diagram, we acquire the free energy.
\alg{
	F_2 = & \int_k \sum_{AB} \psi_{Ak}^*[g^{-1}_{AB} + L_{Ak}\dt_{AB}]\psi_{Bk}.
}
Here, 
\alg{
	g^{-1} =& \f{1}{r_tV_g}[(\ma_0-\ma_x) + r_t\ma_x], \text{ and}\n
	L_{Ak} =& -N_A(S_1(T)-K(T)k^2 + AR(T)hk_x).
}
Also, 
\alg{
	S_1(T) =& \log\f{2e^{\gm_E}E_c}{\pi T} \approx S_1(T_c) -\ep ,\n
	K(T) =& \f{1}{2} S_3(T) \f{(E_R+\mu)}{2m},\n
	R(T) =& \f{1}{2} S_3(T) \sqrt\f{E_R+\mu}{2m},\text{ and}\n
	S_3(T) =& \f{7\zeta(3)}{4\pi^2T^2}.
}
$r_t = \f{2V_u}{V_g+V_u}$, $\gm_E$ is Euler number, $E_c$ is the cutoff energy, $N_A$ is the density of states at band $A$, $\zeta(x)$ is Riemann zeta function, and $E_R = ma^2/2$ is the Rashba energy. 
Since the free energy is a $2\times2$ matrix, we have two distinct eigenvalues. Only one of them is physical for small $r_t$, so we have
\alg{
	F_2' = \int_k \psi_{1k}^* P_1(k) \psi_{1k},
}
where
\alg{
	P_1(k) =&\f{1}{4}(L_{+k}(N_S(k)+N_U(k)) + L_{-k}(N_S(k)-N_U(k)) \nm + 2 N_S(k) S_1(T_c)) 
	+ \f{r_t}{16N_S(k) S_1(T_c)}\nm\times [\{L_{+k}(N_S(k)+N_U(k))+L_{-k}(N_U(k)-N_S(k))\}^2 \nm - 4N_U^2(k) S_1^2(T_c)].
}
Here, $N_S = N_++N_-, N_U = N_+ - N_-$.

Using Eq.~\ref{eq:S33} with $F_2'$, we can obtain the current from paraconductivity for $\mu>0$,
\alg{
	J = \f{e^2}{16\ep}E - \f{ e^3 \pi R(T_c) r_t}{256 S_1(T_c) T_c\ep^{2}} \f{\mu'}{(1+\mu')^{3/2}}h E^2.
}
However, for $\mu<0$,
\alg{
	J = \f{e^2}{16\ep}E.
}
Here, $\mu' = \mu/E_R$. Thus, for $\mu>0$,
\alg{
	\gm = \f{\sqrt{E_R}\pi r_tS_3(T_c)}{2\sqrt2eT_cS_1(T_c)W}\f{\mu'}{1+\mu'},
}
while for $\mu<0$, $\gm =0$. Notably, this model cannot be used to obtain the critical currents, because we employ only up to second order of momentum.

\begin{figure}
	\centering
	\includegraphics[width=\columnwidth]{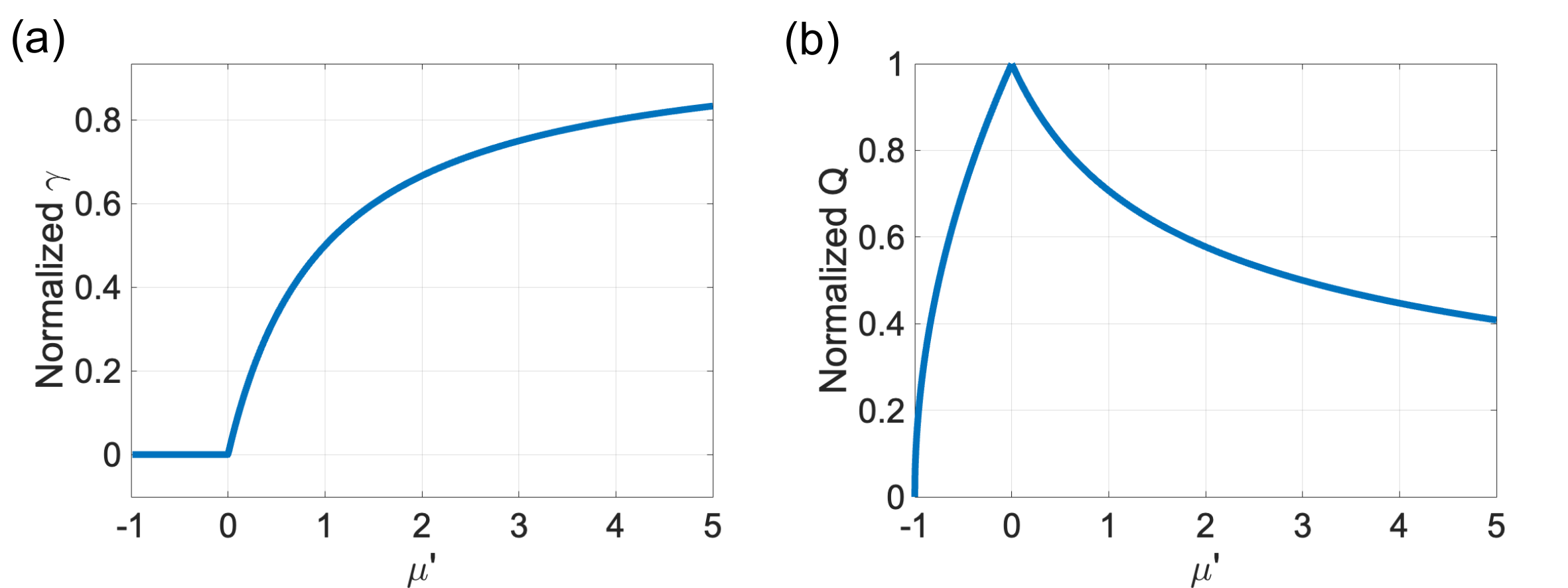}
	\caption{For overdoped case, (a) Normalized $\gm$ and (b) Normalized $Q$ as a function of $\mu'$.}
	\label{fig:S5}
\end{figure}

On the other hand, for $T<T_c$, we only consider $d$-wave interaction and obtain the free energy by bubble diagrams as follows:
\alg{
	F_3 = & \int_k \eta_3(\kb) |\psi_{k}|^2 + \nu_3(\kb) |\psi_{k}|^4. 
}
Here,
\alg{
	\eta_3(k) =&~ \A + \gm_2 k^2 + \gm_4 k^4 + h(\kappa_1 k + \kappa_3 k^3),\n
	\nu_3(k) =&~ \f{1}{2}(\B + h\B_1 k + \B_2 k^2).
}

From the bubble diagram computation, each coefficient is expressed as follows. For $\mu>0$,
\alg{
	\A_+ =& \f{m}{\pi}\ep,\n
	\gm_{2+} =& \f{7\zeta(3)}{16\pi^3} \f{E_R+\mu}{T_c^2},\n
	\kappa_{1+} =& - \f{7\zeta(3)\sqrt{m E_R }}{4\sqrt2 \pi^3 T_c^2},\n
	\kappa_{3+} =& \f{217\zeta(5)\sqrt{E_R}(E_R+\mu)}{128\sqrt{2m} \pi^5 T_c^4}, \n
	\gm_{4+} =& - \f{217\zeta(5)(E_R+\mu)^2}{1024m\pi^5T_c^4},\n
	\B_+ =&\f{21\zeta(3)m}{32\pi^3T_c^2},\n
	\B_{1+} =& \f{279\zeta(5)\sqrt{mE_R}}{64\sqrt2\pi^5T_c^4},\n
	\B_{2+} =& -\f{279\zeta(5)(E_R+\mu)}{256\pi^5T_c^4}.
}
For $\mu<0$,
\alg{
	\A_- =& \f{\A_+}{\sqrt{1+\mu'}},\n
	\gm_{2-} =& \f{7\zeta(3)}{16\pi^3} \f{\sqrt{E_R(E_R+\mu)}}{T_c^2},\n
	\kappa_{1-} =& - \f{7\zeta(3)\sqrt{m (E_R+\mu) }}{4\sqrt2 \pi^3 T_c^2},\n
	\kappa_{3-} =& \f{217\zeta(5)(E_R+\mu)^{3/2}}{128\sqrt{2m} \pi^5 T_c^4}, \n
	\gm_{4-} =& \f{\gm_{4+}}{\sqrt{1+\mu'}}, \nonumber
}
and
\alg{
	\B_- =&\f{\B_+}{\sqrt{1+\mu'}},\n
	\B_{1-} =& \f{279\zeta(5)\sqrt{m(E_R+\mu)}}{64\sqrt2\pi^5T_c^4},\n
	\B_{2-} =& -\f{279\zeta(5)\sqrt{E_R(E_R+\mu)}}{256\pi^5T_c^4}.
}

The qualify factor for $\mu>0$ is
\alg{
	Q = 2.63127\f{h\sqrt{-\ep}}{T_c}\f{1}{\sqrt{1+\mu'}},
}
and for $\mu<0$,
\alg{
	Q = 2.63127\f{h\sqrt{-\ep}}{T_c}\sqrt{1+\mu'},
}
Notably, this model gives zero nonlinear paraconductivity. 

When we apply that $1~T \sim 0.001$, $T_c \sim 0.01$, $D \sim 0.05$, $\chi \sim 1$, $r_t \sim 0.1$, and $J \sim 1.5$, one can get $\gm \sim 10^{5} ~T^{-1}A^{-1}$ and $Q \sim 10^{-1} h\sqrt{-\ep}$. 
Additionally, we present the normalized $\gm$ and $Q$ as a function of $\mu'$ in Fig.~\ref{fig:S5}.

%


\end{document}